\documentclass[10pt,journal,compsoc]{IEEEtran}

\ifCLASSOPTIONcompsoc
  \usepackage[nocompress]{cite}
\else
  \usepackage{cite}
\fi

\usepackage{capt-of}
\usepackage{setspace}
\usepackage{enumitem}
\setlist[itemize]{noitemsep, topsep=0pt}

\usepackage{amsmath}
\usepackage{mathptmx}
\usepackage{graphicx}
\newcommand*\ifigpdfFFTH[1]{\noindent\includegraphics[scale=0.22]{#1}}
\usepackage{fixmath}
\usepackage[LGRgreek]{mathastext}
\usepackage[utf8]{inputenc}
\usepackage{cleveref}
\crefformat{section}{§#2#1#3}
\newcommand*{\uber}{{\it{Uber}}}
\newcommand*{\parsecf}[1]{\textsf{#1}}
\newcommand*{\parsec}{\textsf{PARSEC}}
\newcommand*{\splash}{\textsf{SPLASH}}
\newcommand*{\canneal}{{\parsecf{canneal}}}

\newcommand*{\bodytrack}{{\parsecf{bodytrack}}}

\newcommand*{\facesim}{{\parsecf{facesim}}}

\newcommand*{\fluidanimate}{{\parsecf{fluidanimate}}}

\newcommand*{\streamcluster}{{\parsecf{streamcluster}}}

\newcommand*{\simsmall}{{\parsecf{simsmall}}}

\newcommand*{\simlarge}{{\parsecf{simlarge}}}
\newcommand*{\radix}{{\parsecf{radix}}}
\newcommand*{\ocean}{{\parsecf{ocean}}}
\newcommand*{\SLIP}{{{\it i}{SLIP}}}
\newcommand*{\SLIPone}{1SLIP}
\newcommand*{\SLIPfour}{4SLIP}
\newcommand*\parboldnoin[1]{\vskip 0.1cm \par\noindent {\bf{\fontfamily{phv}\selectfont #1}}}
\renewcommand{\refname}{\footnotesize REFERENCES}

\newcommand*{\numcores}{64-cores}
\newcommand*{\numcorestarget}{256-cores}

\newcommand*{\numcoressmall}{4-cores}
\newcommand*{\numports}{4-port}
\newcommand*{\numportstarget}{16-port}
\newcommand*{\numportstypical}{64-port}
\newcommand*{\numportsrr}{1-port}
\newcommand*{\numportsr}{2-port}
\newcommand*{\numportsl}{8-port}

\newcommand*{\numrouterstages}{4-stage}
\newcommand*{\numlatencytot}{26 cycles}
\newcommand*{\numliftloadlarge}{50\%}                    
\newcommand*{\numliftload}{25\%}                    
\newcommand*{\numliftloadtypical}{2\%}              
\newcommand*{\numliftloadcum}{\numidealloadcum}     
\newcommand*{\numliftqslackfinal}{6$\times$}
\newcommand*{\numliftqslack}{\numlatencyqslack}
\newcommand*{\numliftlatencycrit}{\numlatencycrit}
\newcommand*{\numliftslowdowniq}{\numxbarslowdowniq}
\newcommand*{\numliftperf}{\nummeshperflarge}

\newcommand*{\nummethomsgdata}{72 B}
\newcommand*{\nummethomsgctrl}{8 B}

\newcommand*{\nummethoruntime}{{$\approx$}4 days}
\newcommand*{\nummethoruntimemax}{{$\approx$}10 days}

\newcommand*{\numsetuptotalcapscala}{4 Tb/s}
\newcommand*{\numsetuplinkwidth}{4 B}
\newcommand*{\numsetuplinkcycles}{3 cycles}
\newcommand*{\numsetupcompcycles}{1 cycle}
\newcommand*{\numsetupsercycles}{10 cycles}
\newcommand*{\numidealloadcum}{96\%}
\newcommand*{\numidealload}{0.24 cells/cycle}

\newcommand*{\numidealspikesstart}{10 cycles}
\newcommand*{\numidealspikes}{20 cycles}
\newcommand*{\numidealdelayconc}{18 cycles}
\newcommand*{\numidealdelayswitch}{9 cycles}
\newcommand*{\numidealdelayee}{53 cycles}
\newcommand*{\numidealslowdownr}{1.07}

\newcommand*{\numlatencythre}{40 cycles}
\newcommand*{\numlatencycrit}{20 cycles}

\newcommand*{\numlatencythreslowdown}{1.09}
\newcommand*{\numlatencyqslack}{3$\times$}

\newcommand*{\numxbardelayswitchiq}{51 cycles}
\newcommand*{\numxbarslowdowniq}{1.08}

\newcommand*{\nummeshperflarge}{12\%}
\newcommand*{\nummeshperfsmall}{8\%}
\newcommand*{\formulamisslatency}{
{
\small
\setlength{\abovedisplayskip}{0pt}
\setlength{\belowdisplayskip}{0pt}
\begin{align*}
\text{Miss Latency} &=  L_{L1} + D_{ee}^{ctrl} + L_{L2} + D_{ee}^{data} +         
                       \text{L2 Miss Rate} \times D_{x},
\end{align*}
}
}

\begin{document}

\title{\Large{\bf Uber: Utilizing Buffers to Simplify NoCs for Hundreds-Cores}}

\author{Giorgos Passas \\ gpassas81@gmail.com}

\IEEEtitleabstractindextext{
\begin{abstract}
Approaching ideal wire latency using a network-on-chip (NoC) is an important practical problem for many-core systems, particularly hundreds-cores. Although other researchers have focused on optimizing large meshes, bypassing or speculating router pipelines, or creating more intricate logarithmic topologies, this paper proposes a balanced combination that trades queue buffers for simplicity. Preliminary analysis of nine benchmarks from \parsec\ and \splash\ using execution-driven simulation shows that utilization rises from \numliftloadtypical\ when connecting a single core per mesh port to at least \numliftloadlarge, as slack for delay in concentrator and router queues is around \numliftqslackfinal\ higher compared to the ideal latency of just \numliftlatencycrit. That is, a \numportstarget\ mesh suffices because queueing is the uncommon case for system performance. In this way, the mesh hop count is bounded to three, as load becomes uniform via extended concentration, and ideal latency is approached using conventional four-stage pipelines for the mesh routers together with minor logarithmic edges. A realistic \uber\ is also detailed, featuring the same performance as a \numportstypical\ mesh that employs optimized router pipelines, improving the baseline by \numliftperf. Ongoing work develops techniques to better balance load by tuning the placement of cache blocks, and compares \uber\ with bufferless routing.
\end{abstract}
}

\maketitle

\IEEEdisplaynontitleabstractindextext

\IEEEpeerreviewmaketitle

\IEEEraisesectionheading{\section{Introduction}\label{sec:intro}}

\IEEEPARstart{T}{o} efficiently manage the high wire-to-gate latency ratios in modern VLSI, one popular processor architecture partitions the chip into many identical cores processing in parallel and communicating implicitly via loads and stores to a shared memory that is distributed with the cores \cite{hill, hill_amdhal}. In such systems, a communication medium that uses core-to-core links would imply unmanageable, spaghetti wiring. Thus, together with the processor and memory slice each core also contains a router, and routers are connected in a regular topology, or network-on-chip (NoC). Moreover, NoCs are usually meshes for simplicity. Still, mesh routers are non-negligible overheads, particularly in latency. That being the case, and given the dependence of system performance \cite{smart}, approaching ideal wire latency using a mesh NoC is an important practical problem \cite{smart, mullins_isca, mullins_asia, expressvcs_isca, expressvcs_dac}. Moreover, this problem becomes compounded as systems scale \cite{nilmini, smart}. Solutions include techniques to bypass routers \cite{smart, expressvcs_isca, expressvcs_dac} and low-latency router design \cite{mullins_isca, mullins_asia}. However, such techniques increase design complexity, while router overheads might remain high. What is more, wide links and high frequencies for fast serialization of core messages have led to such low utilizations \cite{circuitcoherence, cambridge, gratz, hesse} that bufferless routing has been defended \cite{bless}. Taking into consideration all the above data, this paper proposes to compensate for system scaling by extending core concentration. Although delays in concentrator and router queues increase as mesh port load rises, queueing is the uncommon case for system performance. Besides, concentrators are local structures, and more intricate, globally logarithmic topologies \cite{balfour} are obviated. In this way, this paper makes a clear case for a buffered NoC.

In particular, from present small systems to hundreds-cores of the near research future, mesh hop count makes a critical step, while port utilization does not exceed a typical low of \numliftloadtypical. Concentrating few cores is usual \cite{balfour}, but non scalable. This paper extends concentration to 16 and beyond, thus increasing utilization to \numliftload\ or higher. The utilization boost is enabled by state-of-the-art benchmarks that tolerate queue delays at least up to \numliftqslack\ higher compared to an ideal latency of just \numliftlatencycrit. Owing to this slack, a \numportstarget\ mesh scales to hundreds-cores, and hop count is easily bounded to three. Overall, the traditional \numrouterstages\ pipeline \cite{ppin} suffices without any bypassing or speculation optimizations, and buffers implementing queues play a key role in the simplification. Although some researchers have already studied evenly-utilized configurations \cite{carpenter, hesse}, or even similar topologies \cite{hybrid}, their analysis focuses on small systems or custom interconnects, hence missing the role of buffers.

Compatibly with previous studies (e.g. \cite{hesse}), numbers correspond to analysis of seven benchmarks from \parsec\ \cite{parsec} and two benchmarks from \splash\ \cite{splash} on \numcores\ in the gem5 full-system simulator \cite{gem5}. In this context, this paper measures a cumulative load of \numliftloadcum\ on average, and spreads this load to four ports. Assuming that (i) the need for bandwidth grows linearly with system size \cite{nilmini}, and (ii) benchmarks for hundreds-core should demonstrate similar communication patterns like \parsec, such a \numports\ mesh is miniature of a \numportstarget\ counterpart in \numcorestarget. To better comprehend stress, the analysis abstracts the mesh using a single-stage switch that implements ideal output queueing \cite{karol}. As a transient step to reality, output queueing is replaced first by models of idealistic single-stage crossbars \cite{karol, passas}. In comparison, using a single FIFO per input, queue delay slips by head-of-line (HoL) blocking \cite{karol} slightly beyond the nominal slack, resulting to tangible system slowdown that measures to \numliftslowdowniq. Using more advanced organizations \cite{mckeown}, queueing is sufficiently bounded, while scheduling efficiency plays a marginal role. Furthermore, because extended concentration makes the load uniform and bursty \cite{poisson_floyd} crossbars are redundant, and are replaced by a mesh. Performance remains excellent, owing to router buffers that provide a kind of speedup. Comparing a \numports\ and a \numportstypical\ mesh, both using \numrouterstages\ routers, the small instance reduces performance loss from \nummeshperflarge\ to \nummeshperfsmall, although it presents higher end delay by queueing, and falls short of a large instance that offers ideal performance using single-cycle routers. This handicap is attributed to uncontrolled interleaving of control and data cells. Indeed, a more realistic organization that separates messages in virtual networks removes the above handicap. What is more, virtualization roughly doubles the slack for queueing. 
\clearpage
\parboldnoin{My Contributions:}
\begin{itemize}
\item Analysis to compare the latency and queue delay impact of a NoC on system performance
(Sec. \ref{sec:ideal}, Sec. \ref{sec:latency})
\item Analysis to plot HoL blocking as a side-effect of utilization boost
(Sec. \ref{sec:ideal}, Sec. \ref{sec:xbar})
\item A highly utilized \numportstarget\ mesh for hundreds-cores featuring simpler pipelines by better 
utilized buffers 
(Sec. \ref{sec:mesh}, Sec. \ref{sec:uber})
\end{itemize}

\vspace{-0.3cm}
\section{System and NoC Model}

\label{sec:setup}

The focus is on a concrete memory organization, similar as the baseline in \cite{hill}. Memory blocks are \nummethomsgdata\ and coherence control messages are \nummethomsgctrl. More details are given in Table 1. With respect to benchmarks, this paper considers \parsec\ \cite{parsec} and \splash\ \cite{splash} for comparison with \parsec. CPI is on average seven, when measured on \numcoressmall. For experiments using unloaded NoCs, I was always able to provide five runs for all benchmarks except for \streamcluster\ (one run) and \facesim\ (two runs). With very stressed NoCs, runs are in general fewer. The critical path is \nummethoruntimemax\ and the typical case is \nummethoruntime. Finally, for \parsec, input sets are \simlarge, whereas for \splash\, inputs correspond to \simsmall\ sets.
 
\begin{figure}[t]
	\begin{tabular}{cc}
	\begin{minipage}[t]{.28\textwidth}
		\centering
		\vspace{0pt}
		\captionof{table}{Additional System Parameters}
{\footnotesize
		\begin{tabular}{|r||l|}
			\hline
			PROCESSOR & 2GHz, in order					 \\
			\hline
			L1 CACHE  & (64+32)KB, private,              \\
    	              & 2 ways, 3 cycles                 \\
			\hline
			L2 CACHE  & 2MB/core \cite{bless}, shared,	\\
			          & 8 ways, 12 cycles,			    \\
					  & MESI full-map dir				\\
			\hline
			MEMORY    & 100 cycles,                     \\
			          & 72B blocks						\\
			\hline
			\splash   & 258$\times$258 matrix,			\\
				            & 1M integers               \\
			\hline
			OS        & Linux 2.6.27,                 	\\
			          &               1 thread/core		\\
			\hline
		\end{tabular}
}
	\end{minipage}
	&
	\begin{minipage}[t]{.2\textwidth}
		\centering
		\vspace{32pt}
		\ifigpdfFFTH{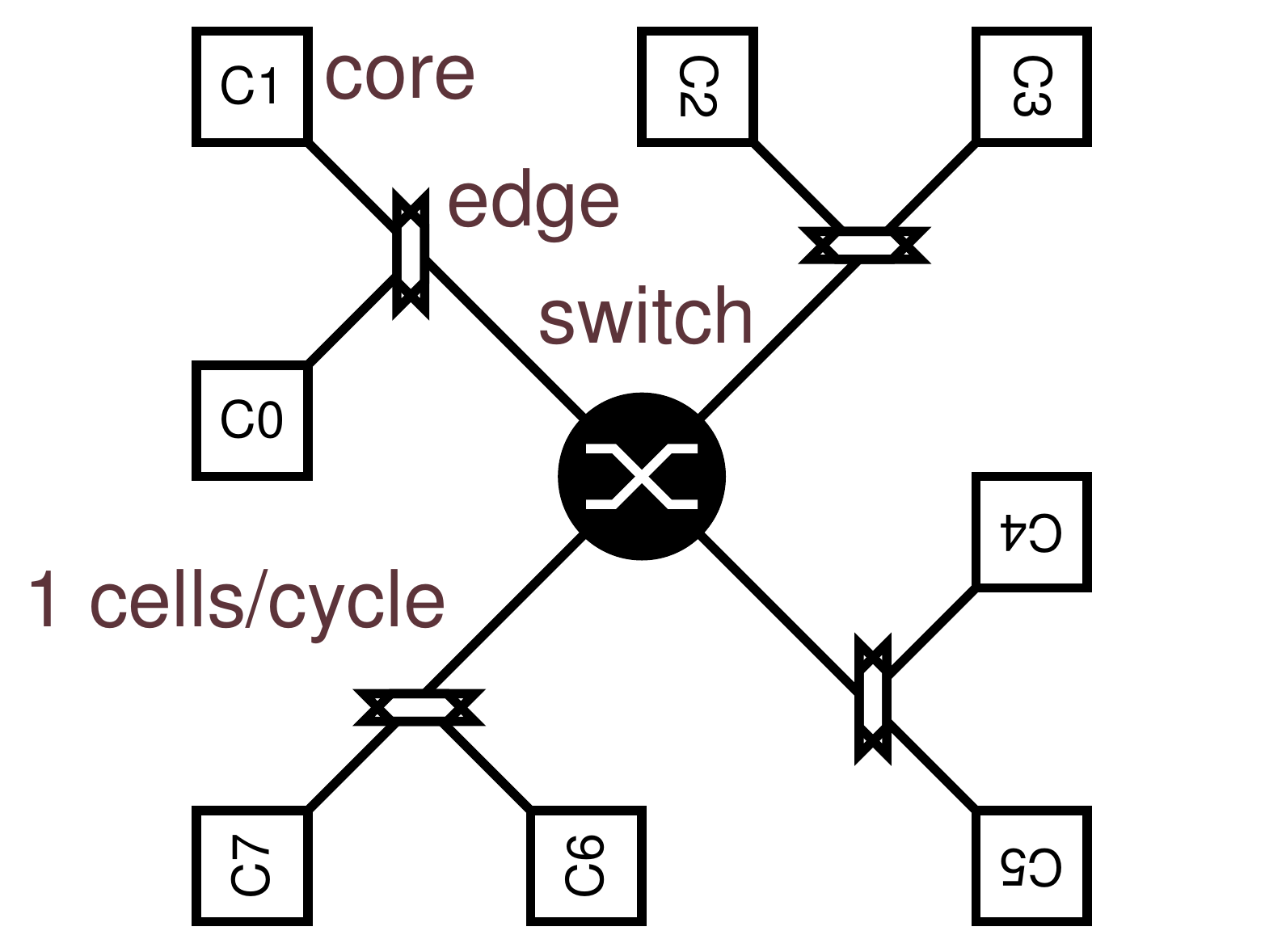}
		\captionof{figure}{NoC model diagram}
	\end{minipage}
	\end{tabular}
\end{figure}

Fig. 1 gives a diagram of the NoC. Core messages are fragmented into fixed-size cells, which cells are injected through edge multiplexors, switched, and ejected through the demultiplexors. Once messages are reassembled from cells at their destination core, they are delivered to the correct controller. The term {\it cell} emphasizes that queues are primarily infinite. In particular, the switch employs one queue per output so that cells that arrive concurrently at the switch inputs are written in parallel and indiscriminately. Such output queueing \cite{karol} is used widely for ideal-performance switching. Output queueing is also useful to model random multiplexing at the edges. Note that contending cells are always downstream, and there is no contention from edges to cores. Following the above discussion, although the switch is ideal, the whole NoC is not. Referring to Fig. 1, consider three cells at cores C0, C1, and C2 destined to C4, C6, and C6, respectively (C6 is double to denote contention). When C1 is ``accidentally'' prioritized over C0 at the edge, the NoC suffers one extra delay cycle. Nonetheless, I was unable to plot any performance loss by such edge blocking. The pipeline comprises \numsetuplinkcycles\ at the links, plus \numsetupcompcycles\ at every other component (including the cores). By default links are \numsetuplinkwidth\ instead of a typical 16 B \cite{hesse} to reserve about 4$\times$ bandwidth for faster systems and more demanding benchmarks. Thus, adding \numsetupsercycles\ for message serialization, the total latency is \numlatencytot. End delay is offset by queueing and message reassembly. 

\begin{figure}[t!]
	\centering
	\begin{tabular}{lll}
		\ifigpdfFFTH{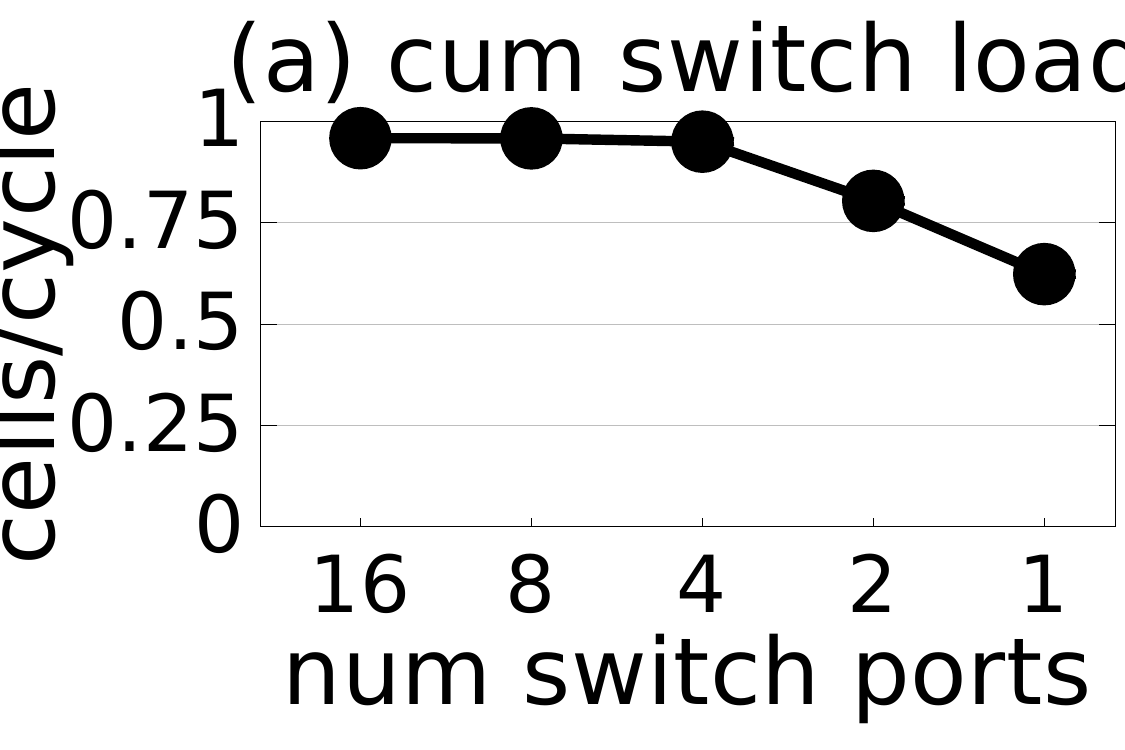}
		&
	    \ifigpdfFFTH{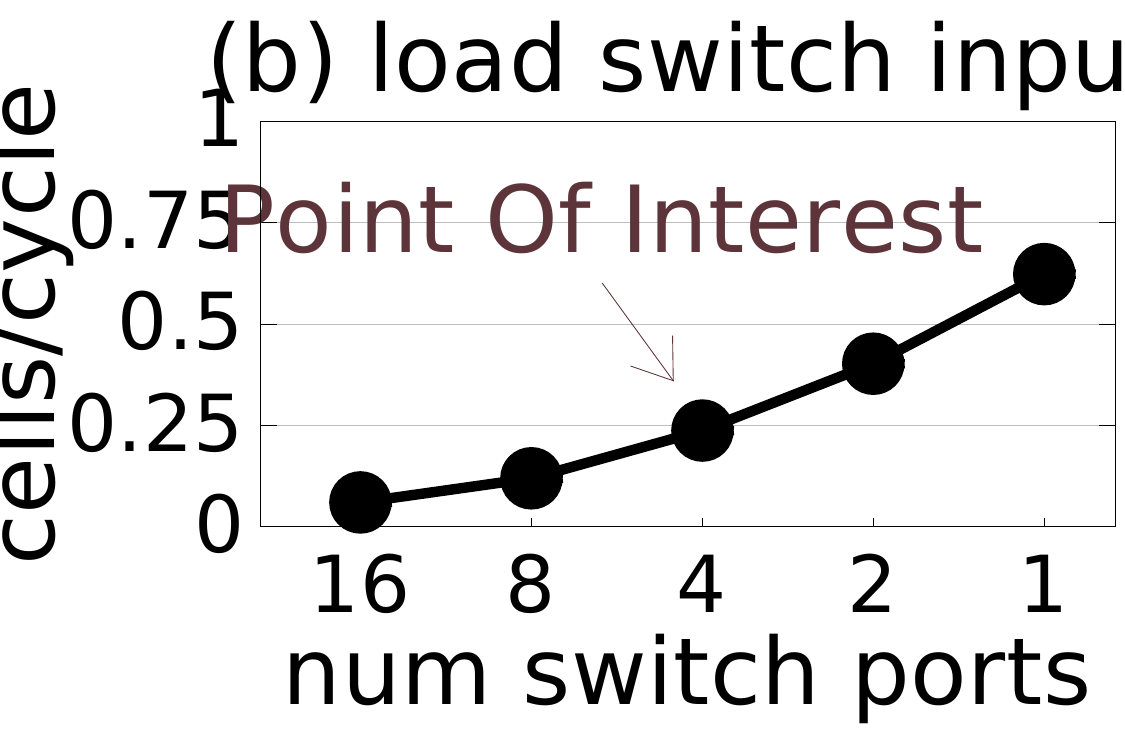}
		& 
		\ifigpdfFFTH{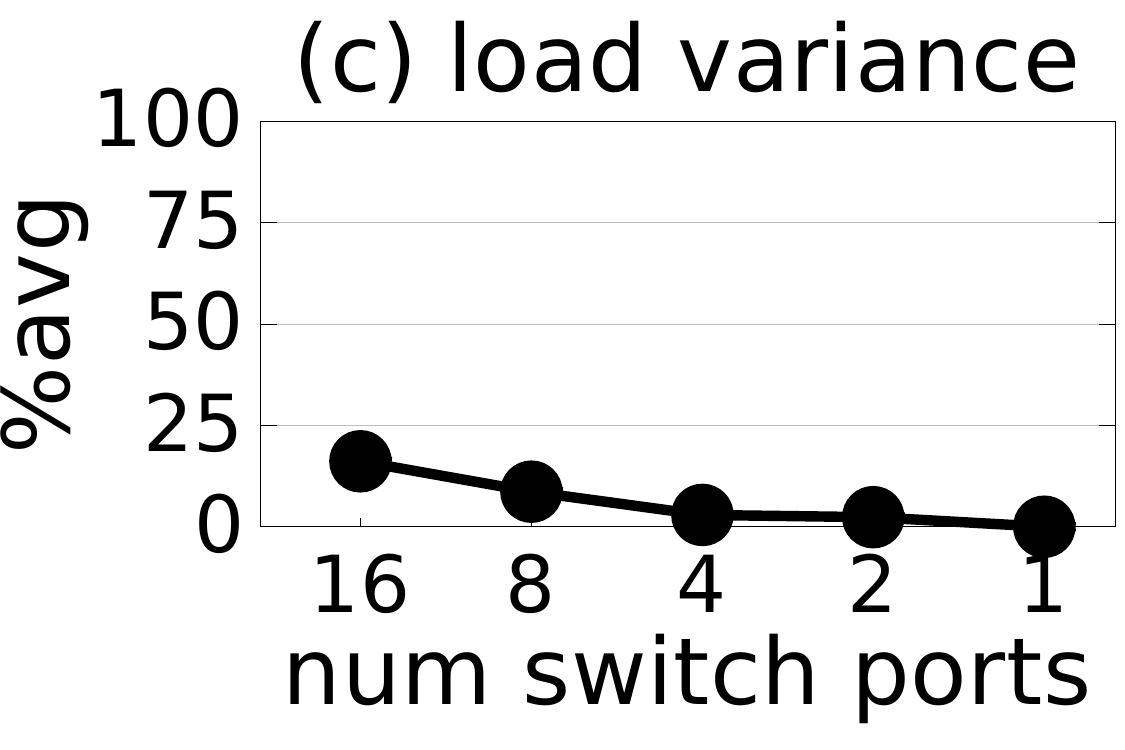}
		\\
		\ifigpdfFFTH{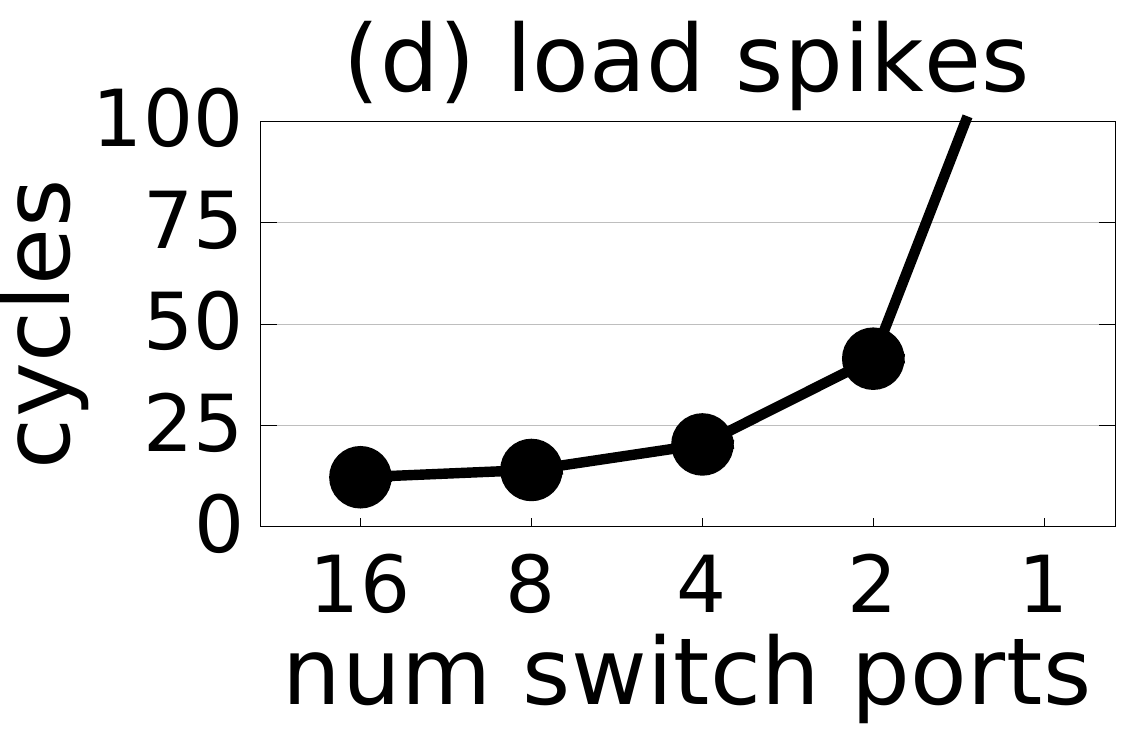}
		& 
	    \ifigpdfFFTH{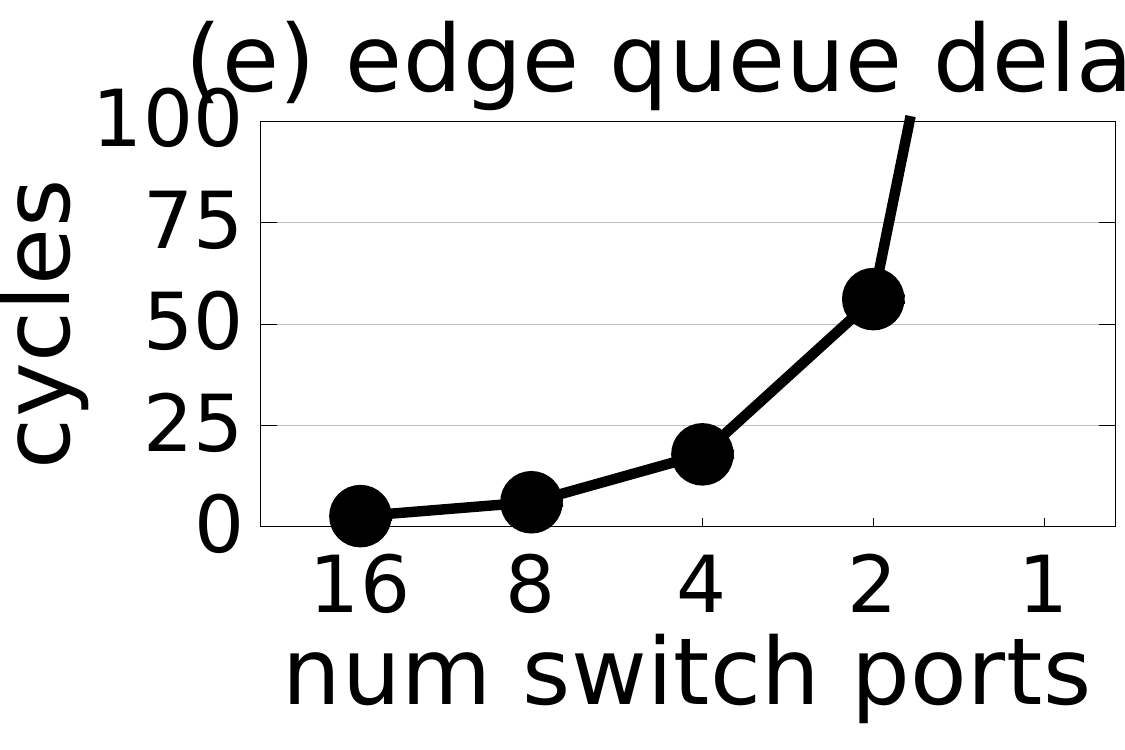}
		&
		\ifigpdfFFTH{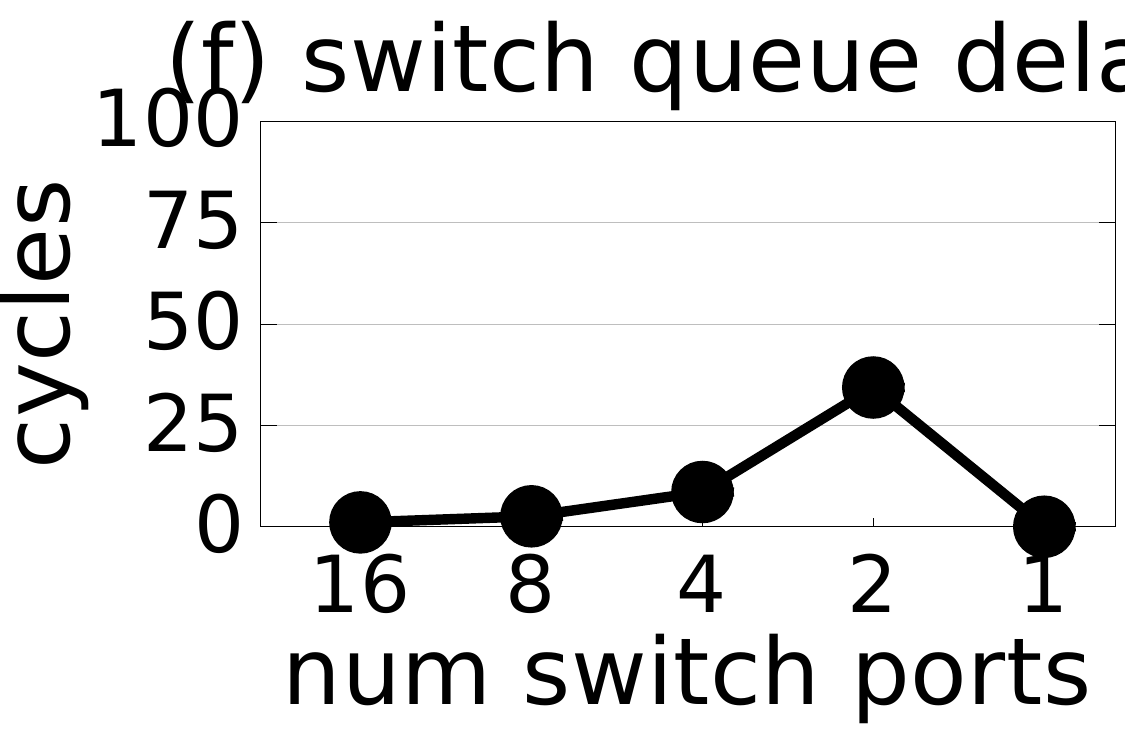}
		\\
		\ifigpdfFFTH{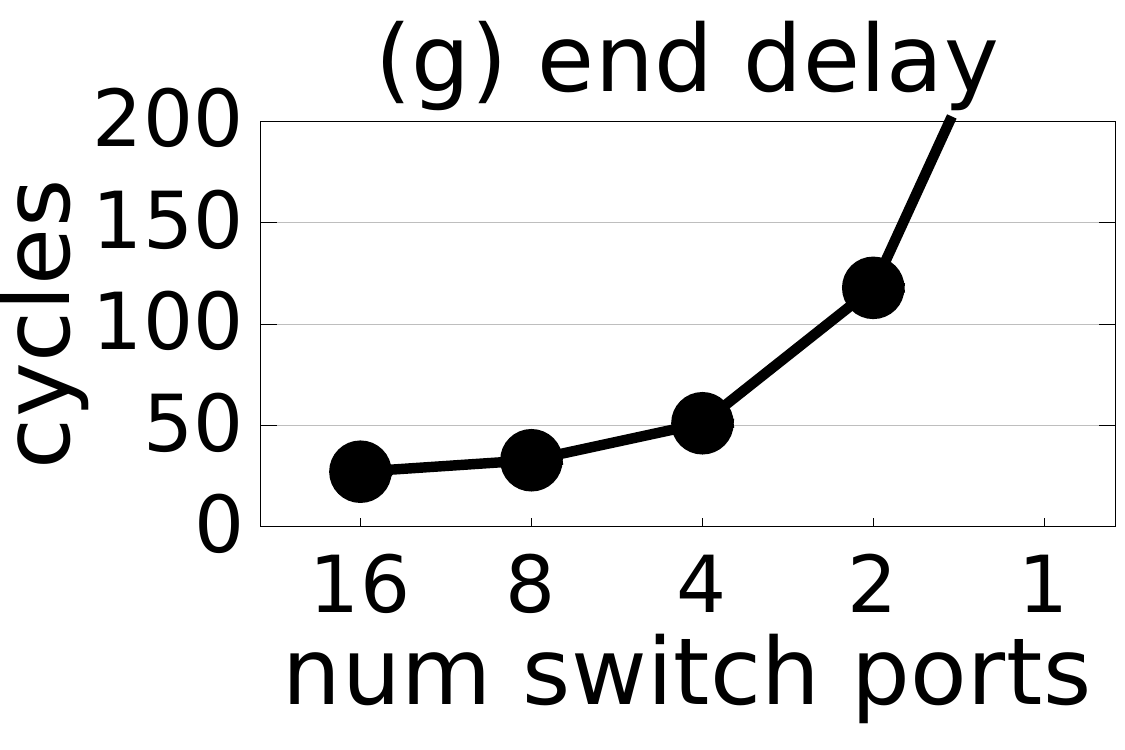}
		& 
		\ifigpdfFFTH{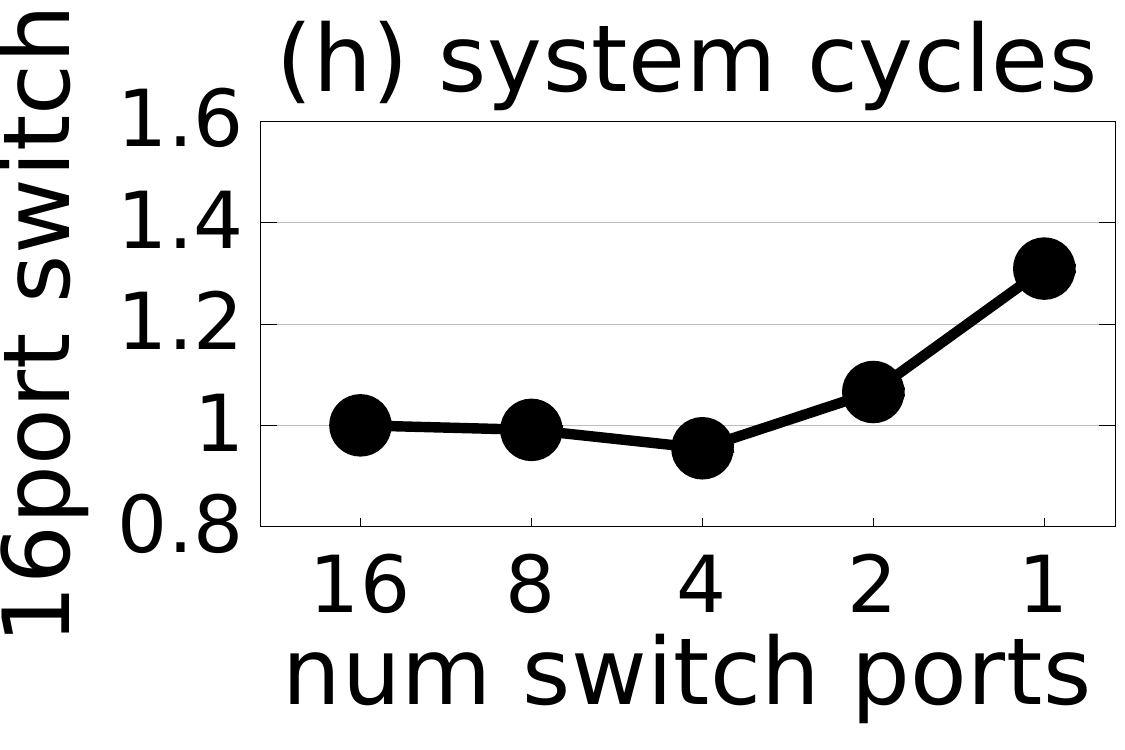}
		&
		\ifigpdfFFTH{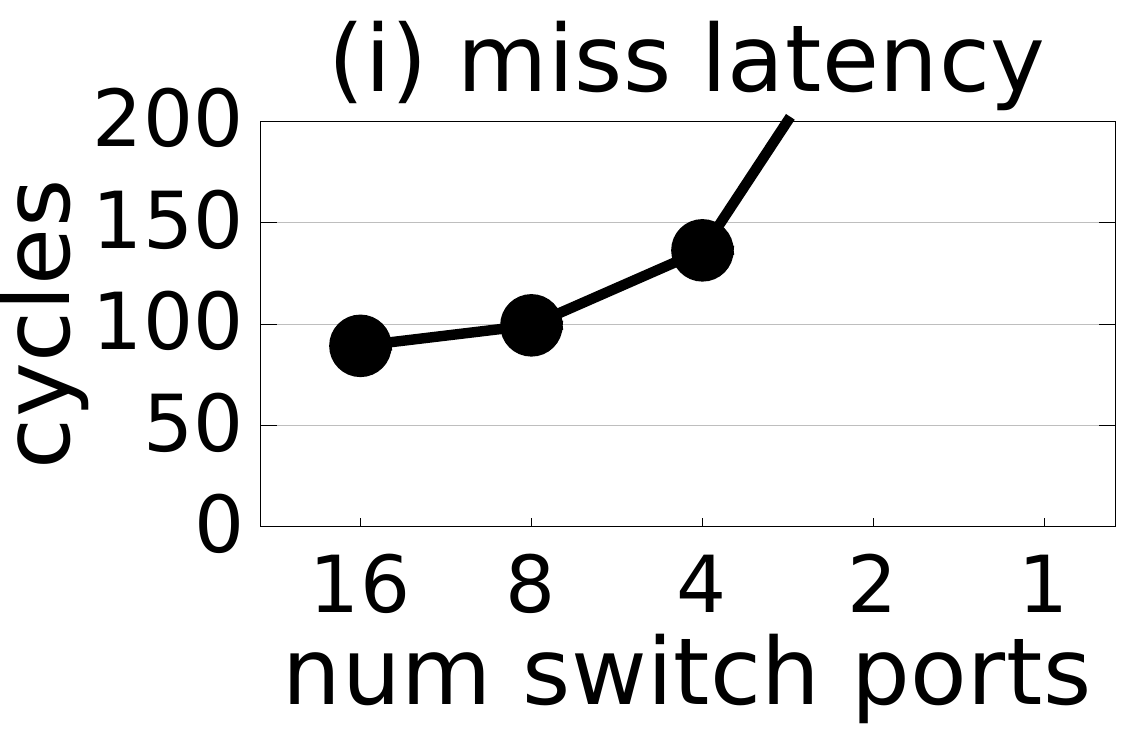}
	\end{tabular}
    \caption{
            (a-g) NoC stress and (h-i) system performance;
			each point is the average of nine benchmarks
            }
    \label{figs:gnuplot:bot}
\end{figure}
\section{Evaluations Using Idealistic NoCs}

\label{sec:ideal}

This section evaluates NoC stress and the impact thereof on system performance, focusing on \numports\ switches in \numcores\ as miniatures of \numportstarget\ switches for \numcorestarget. The main variable is the size of the switch, which ranges from 16 ports down to one port. A NoC using a \numportsrr\ switch is actually a bus \cite{carpenter, udipi}.

Thus, Fig. \ref{figs:gnuplot:bot} plots nine metrics averaged over the nine benchmarks of \parsec\ and \splash. In (a), only a single link is fully utilized on average. Such a low utilization is well known in the literature \cite{circuitcoherence, cambridge, gratz, hesse}. Moreover, load drops using small switches. System slowdown largely explains this throttling. In (b), cumulative load is averaged over the switch inputs. We observe that load remains below 1 cells/cycle. The number of interest is \numidealload\ for \numports\ switches. In (c), switch inputs are more uniformly loaded for smaller switches. In such a way, edges never saturate. Indeed, edge queueing is plotted comparatively short below. A similar discussion applies also for the outputs of the switch, hence neither does the switch saturate. In (d), we observe that switch load is spiked. Spikes start from \numidealspikesstart\ ---corresponding to roughly one control message every other cache block--- growing up to several tens of cells. For \numports\ switches, spikes are \numidealspikes. Although NoC load patterns are known to contain spikes \cite{cambridge, gratz}, this paper is from the first to clearly plot their impact on system performance, in a way merging spikes in time and/or space. Overall, switch loading resembles the Internet core \cite{poisson_floyd}. Load metrics are helpful for understanding NoC behavior, but what matters for the system is end delay. Fig. \ref{figs:gnuplot:bot}(e) and (f) contribute queue delays at the edges and the switch, respectively. We observe that queue delays increase as the switch shrinks. Using \numports\ switches, \numidealdelayconc\ are spent at the edges and \numidealdelayswitch\ at the switch. This 2$\times$ ratio is consistent also for smaller switches. Further than queue delays, the time it takes to reassemble messages from cells at the cores increases. In any case reassembly delay is almost negligible (not shown). Next, (g) plots end message delay. Using \numports\ switches, end delay is on average \numidealdelayee, of which \numlatencytot\ is the latency term and the remainder is the sum of queue delays at the edges and the switch. Note, we assumed that data and control messages suffer equal delays in the NoC. This paper found the above assumption true, owing to NoC multiplexing being random. Finally, the impact of queue delays on system performance is plotted in Fig. \ref{figs:gnuplot:bot}(h). Shrinking the switch down to \numports\ is no trouble. Slowdown starts using \numportsr\ switches; and measures to \numidealslowdownr\ in this case. Next, Fig. \ref{figs:gnuplot:bot}(i) plots the L1 miss latency, as reported in the system statistics \cite{gem5}. Using this first-order approximation,
\formulamisslatency
hits in L2 explain 75\% of the miss latency. 

\section{Queue Delay Compared to Latency}

\label{sec:latency}

\begin{figure}[t!]
	\begin{tabular}{cc}
		\ifigpdfFFTH{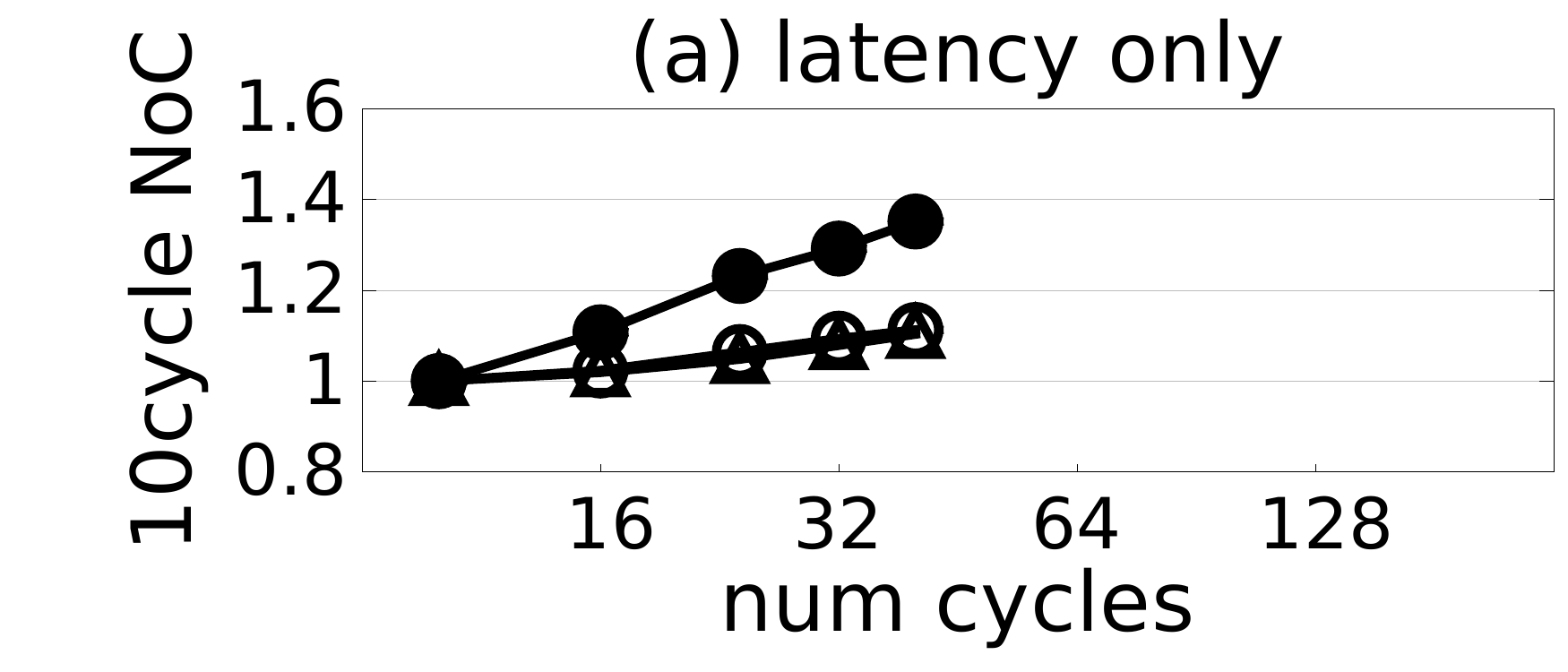}
		&
		\ifigpdfFFTH{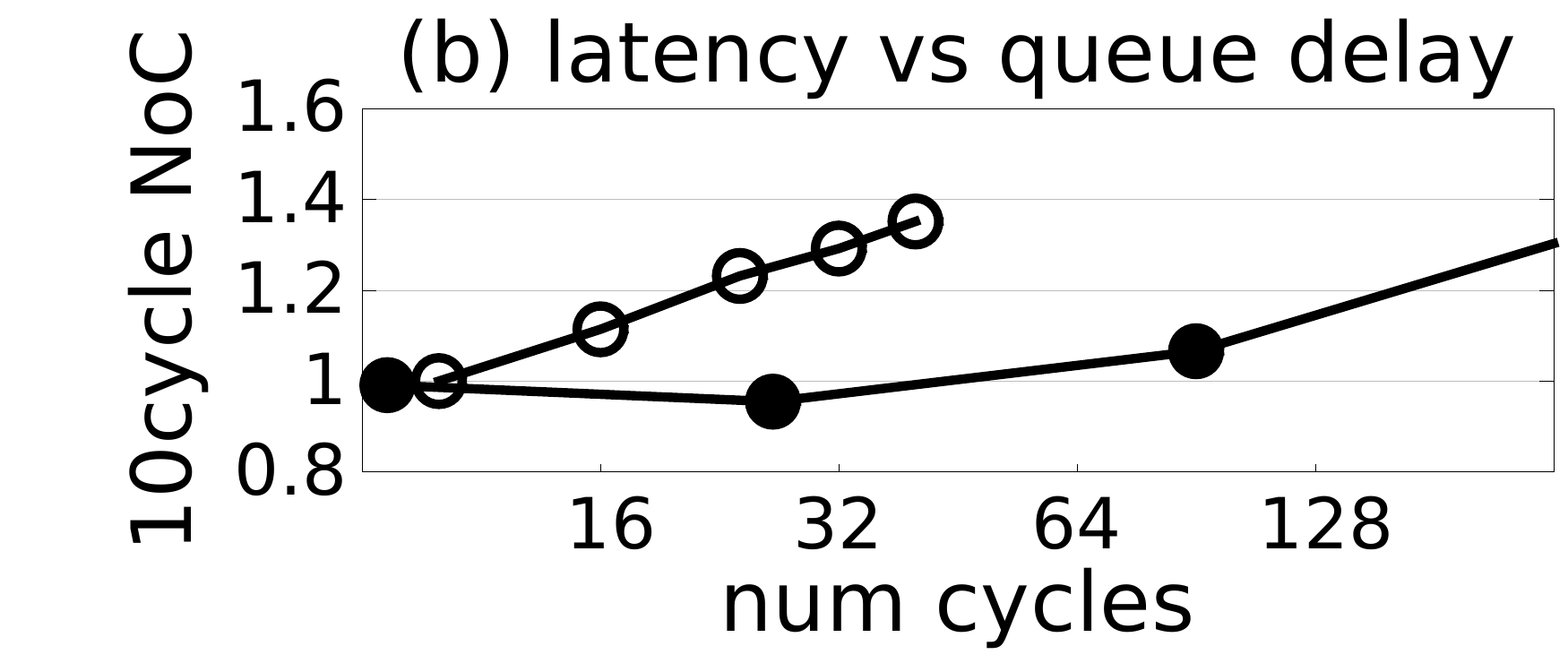}
		\\
		\ifigpdfFFTH{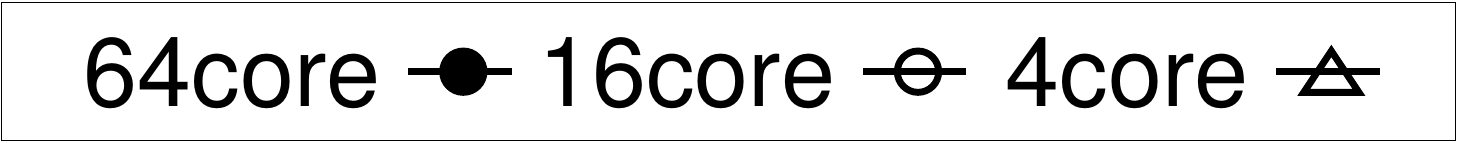}
		&
		\ifigpdfFFTH{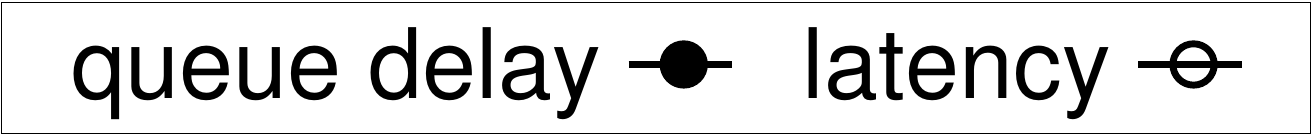}
	\end{tabular}
    \caption{
			Effect of latency and queue delay on system performance;
			latency is total, end-to-end; queue delay is total, at the edges and the switch
            }
    \label{figs:gnuplot:ll}
\end{figure}

This section adds results from a complementary experiment that varies NoC latency by explicitly offsetting the links. Fig. \ref{figs:gnuplot:ll}(a) plots system performance under ranging NoC latency for three distinct system sizes. We observe that \numcores\ are more sensitive than smaller systems. This differentiation is due to a single benchmark, namely \streamcluster. Either way, \numlatencythre\ imply system slowdown of at least \numlatencythreslowdown. Though marginal, slowdown is clear even at \numliftlatencycrit. Although Krishna et al. \cite{smart} measure significant losses at even lower latencies, Enright-Jerger et al. \cite{circuitcoherence} are more conservative, as is this paper. Fig. \ref{figs:gnuplot:ll}(b) compares latency and queue delay. The queue delay curve merges the sum of edge and switch delays in Fig. \ref{figs:gnuplot:bot}(e) and (f) with the corresponding system performance in Fig. \ref{figs:gnuplot:bot}(h). Thus, each data point refers to NoC using a distinct switch instance from \numportsl\ down to \numportsrr. In comparison, queue delay can grow \numliftqslack\ higher. This experimental result is intuitive: both inside each individual benchmark and across the whole set, communication can be seen in phases where load is typically low \cite{cambridge}.

\section{Replacing Ideal Switches with Crossbars}

\label{sec:xbar}

This section evaluates system performance degradation when ideal output queueing is replaced with more practical crossbars. This transition is particularly interesting given recent studies that have demonstrated the feasibility of large crossbars on chip \cite{passas}. 

\begin{figure}[t!]
	\begin{tabular}{c}
	\begin{tabular}{cc}
		\ifigpdfFFTH{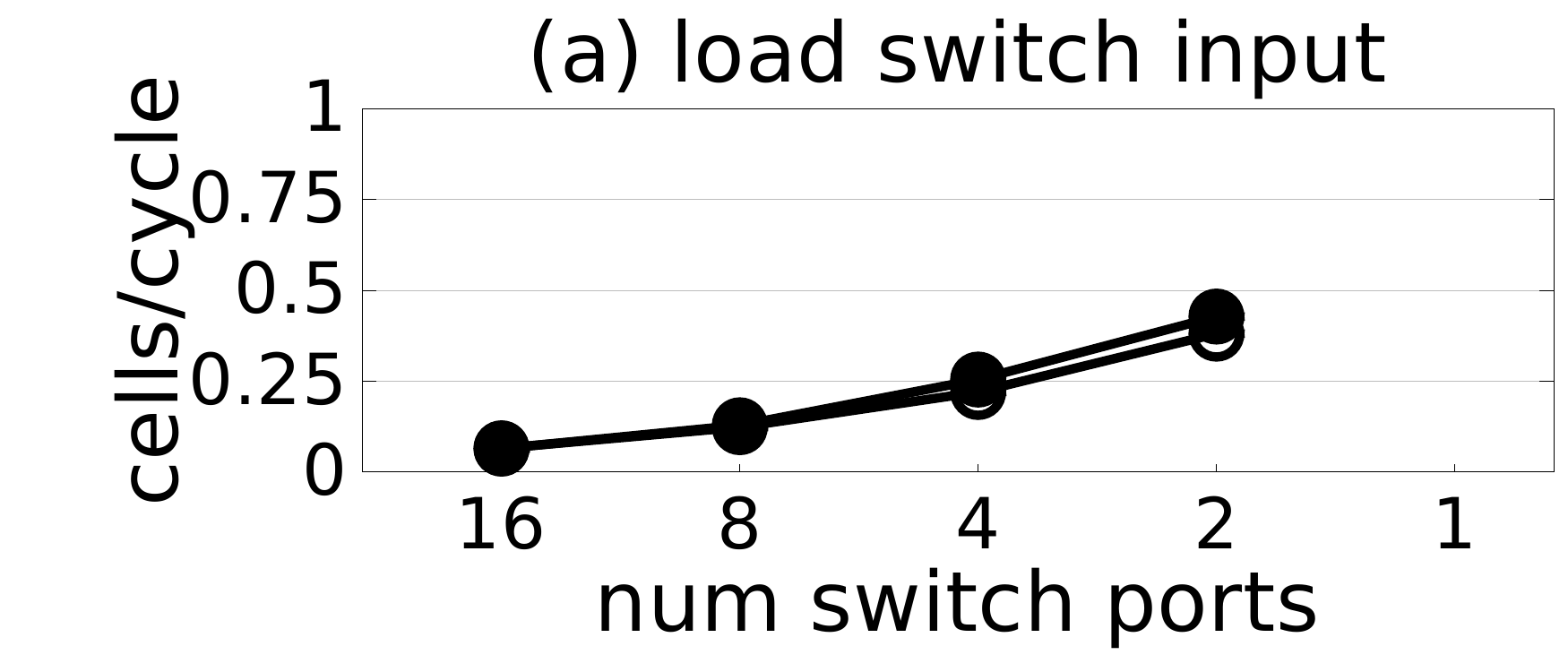}
		&
		\ifigpdfFFTH{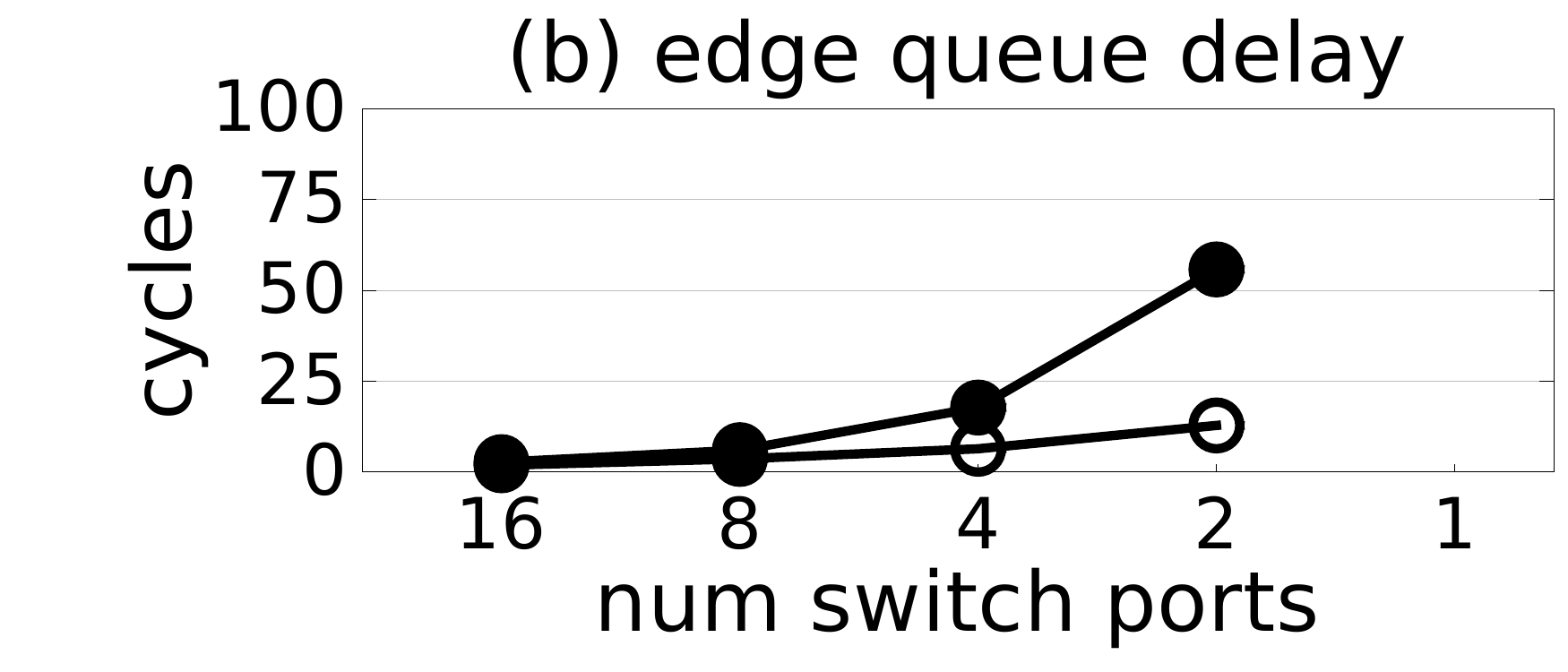}
		\\
		\ifigpdfFFTH{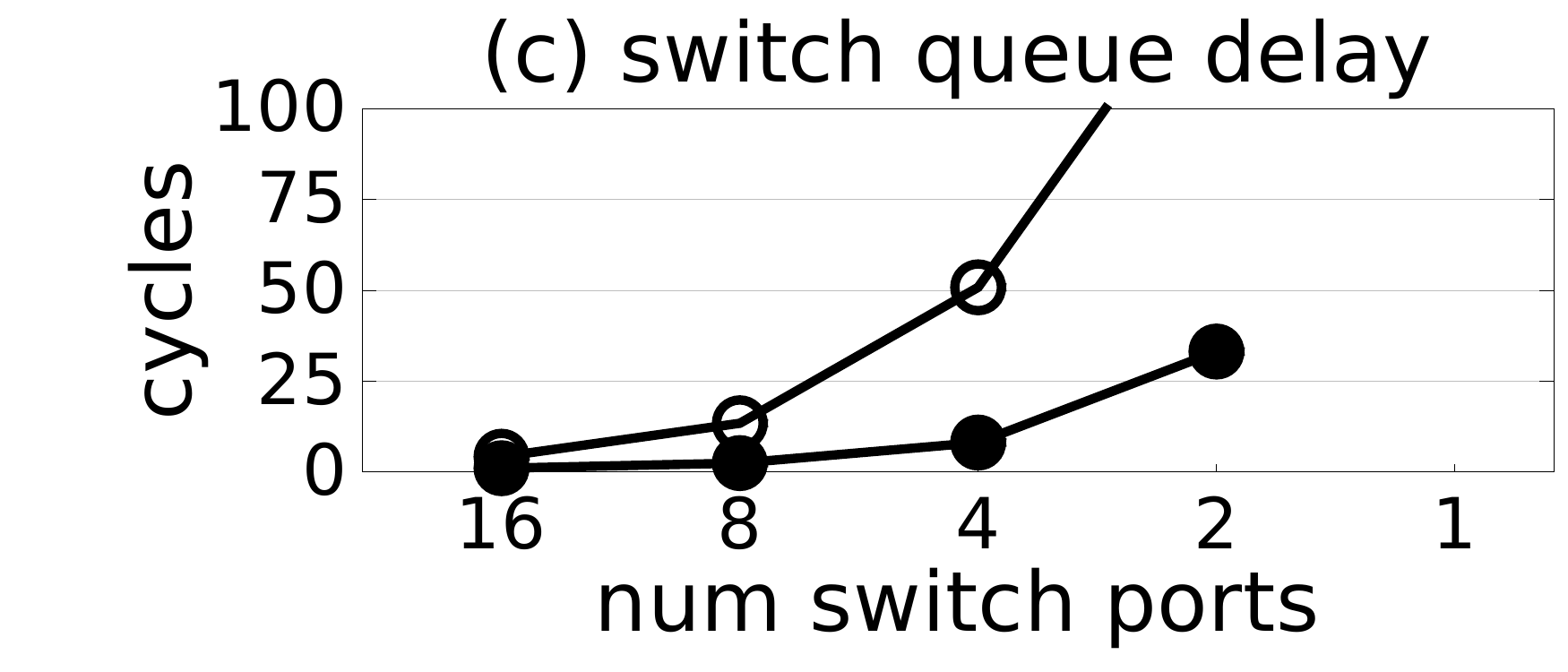}
		&
		\ifigpdfFFTH{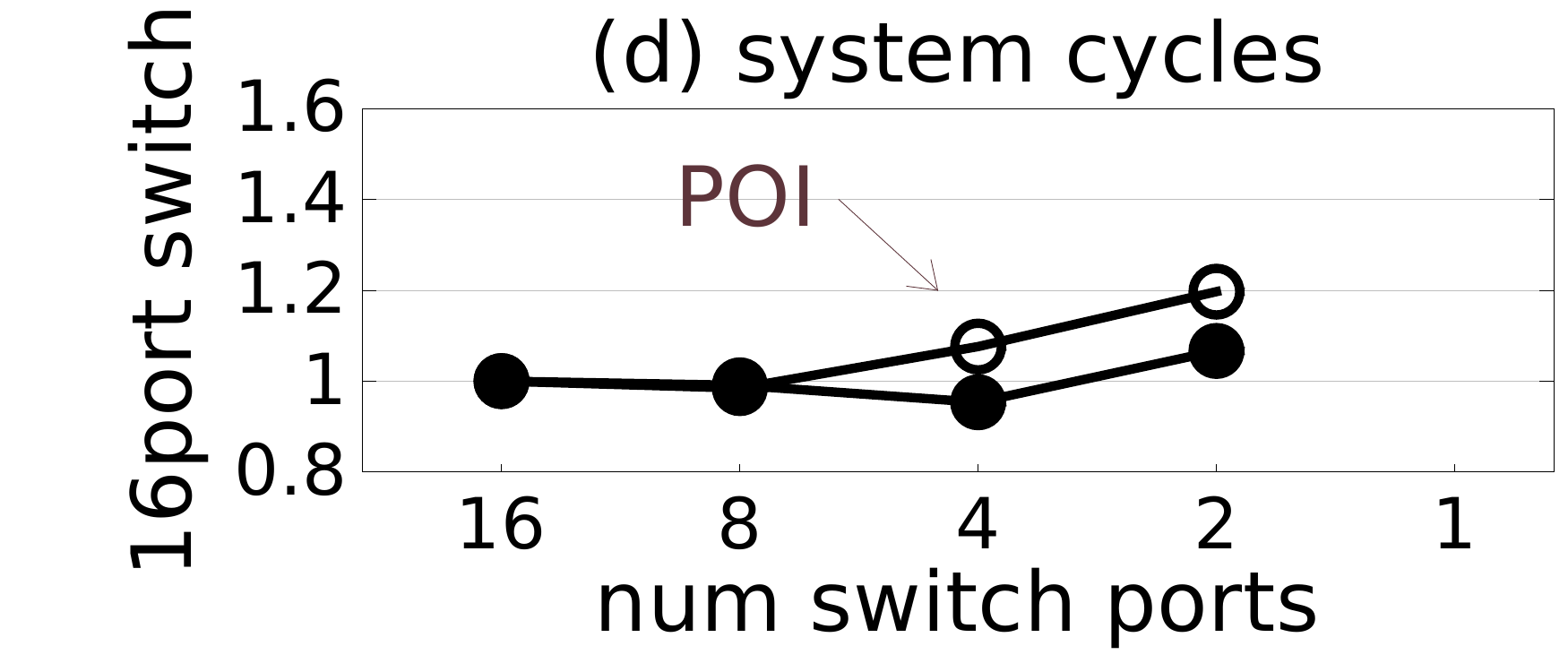}
		\end{tabular}
		\\
		\ifigpdfFFTH{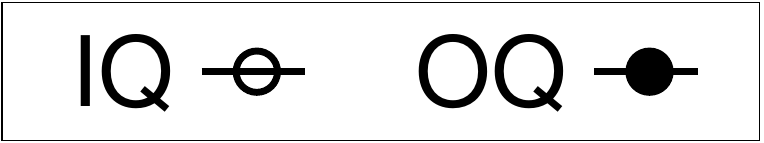}
	\end{tabular}
    \caption{
			Effect of HoL blocking on NoC and system performance
			}
    \label{figs:gnuplot:xbar_iq}
\end{figure}

A first model of input queueing (IQ) employs a single FIFO queue per input together with one round-robin arbiter per crossbar output \cite{karol}. Fig. \ref{figs:gnuplot:xbar_iq} compares IQ with OQ. We observe in (a) that switch load is slightly lower in IQ. Edges reach equilibrium with the switch as benchmarks slow down by HoL blocking \cite{karol} at the switch, despite the fact that there is no backpressure (unlimited queues). In turn, edge queueing is also shorter as in Fig. \ref{figs:gnuplot:xbar_iq}(b). However, switch queueing is longer by HoL blocking as in (c). For \numports\ switches, delay grows from \numidealdelayswitch\ in OQ to \numxbardelayswitchiq\ in IQ. Thus, delay is longer overall, and benchmarks slow down as in Fig. \ref{figs:gnuplot:xbar_iq}(d). For \numports\ switches, slowdown measures to \numxbarslowdowniq. Note that the above delay-performance relation is a good match with the analysis in Sec. \ref{sec:latency}. 

\begin{figure}[t!]
	\centering
	\begin{tabular}{c}
		\ifigpdfFFTH{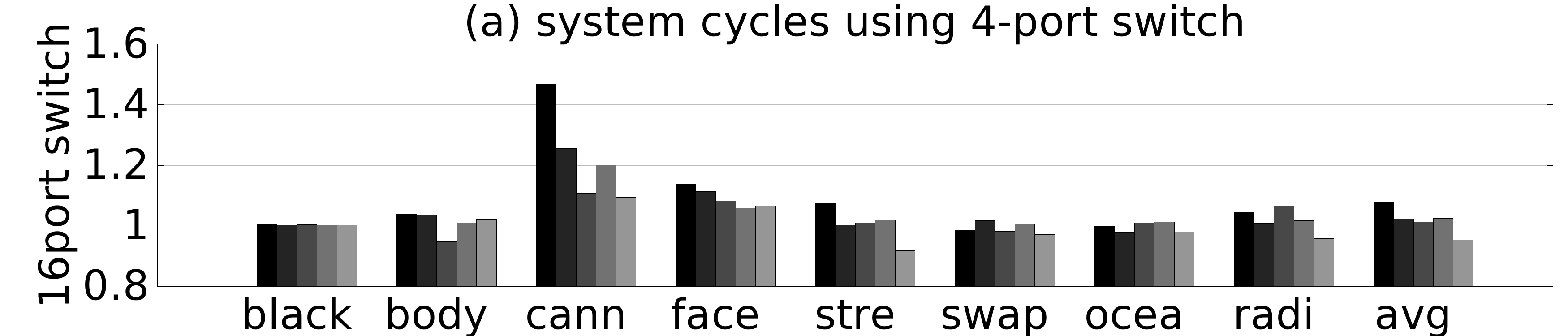}
		\\
		\ifigpdfFFTH{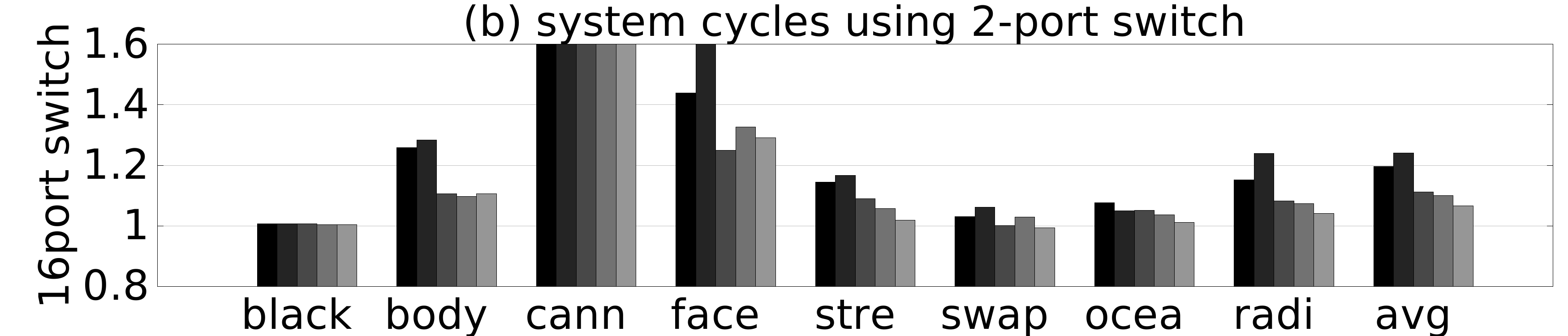}
		\\
		\ifigpdfFFTH{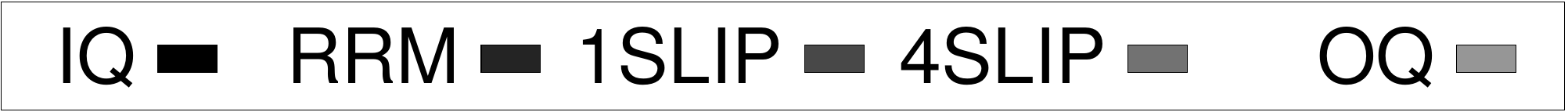}
	\end{tabular}
    \caption{
			Crossbar performance per benchmark; \fluidanimate\ failed for some configurations
			}
    \label{figs:gnuplot:xbar_bench}
\end{figure}

There is a large body of research on improving IQ. A popular solution is Virtual Output Queueing together with a crossbar scheduler like \SLIP\ \cite{mckeown}. Fig. \ref{figs:gnuplot:xbar_bench} plots system performance separately for each benchmark for \numports\ and \numportsr\ switches. We observe that \bodytrack, \canneal, and \facesim\ are critical benchmarks that differentiate scheduling performance. On average, however, differences are impractical. Moreover, observe in Fig. \ref{figs:gnuplot:xbar_bench} that there is an anomaly in the performance of \canneal\ with \SLIPfour, despite the fact that \SLIPfour\ queueing is actually as good on average. This anomaly returns in Sec. \ref{sec:mesh} for the mesh. The corner behavior of RRM \cite{mckeown}, on the other hand, is well explained by synthetic-load experiments. In conclusion, HoL blocking and its efficient reduction is a basic problem associated with utilization boost. Either way, the role of techniques to resolve HOL blocking is corrective, to prevent queueing from slipping beyond a nominal slack. 

\section{Replacing Crossbars with a Mesh}

\label{sec:mesh}

This section details the transition from crossbar to mesh. A mesh provides a finer physical hierarchy and also simplifies queueing. Parameters of interest are as follows: Routers are always dimension-ordered, router queues infinite FIFOs at inputs, router latency \numrouterstages, and link latency one cycle. For small instances, mesh hop count is bounded to three, hence latency is as for the crossbar. For \numportstypical\ instances, however, hop count jumps to seven, and latency makes a critical step. 

\begin{figure}[t!]
	\centering
	\begin{tabular}{c}
	\begin{tabular}{cc}
		\ifigpdfFFTH{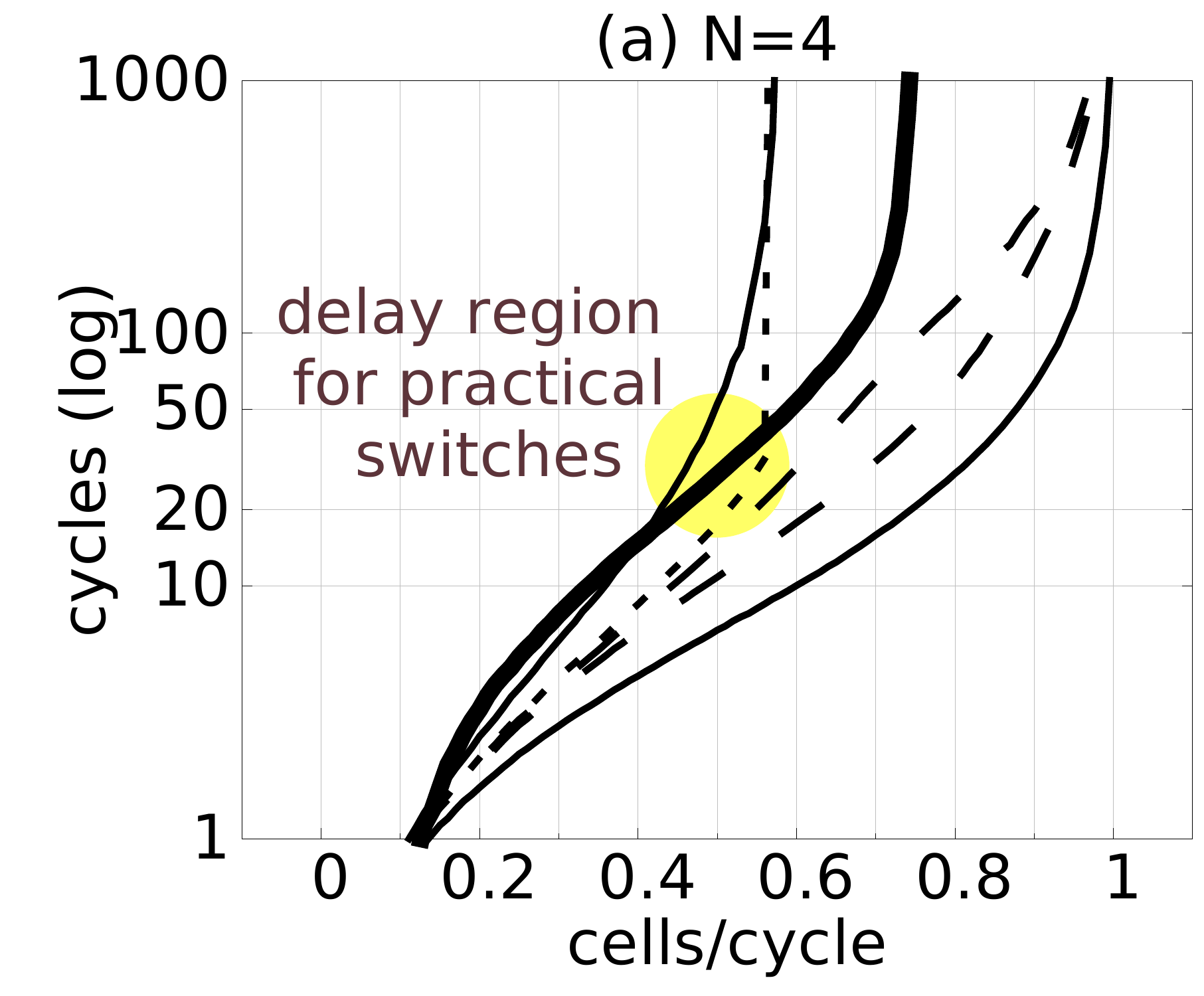}
		&
		\ifigpdfFFTH{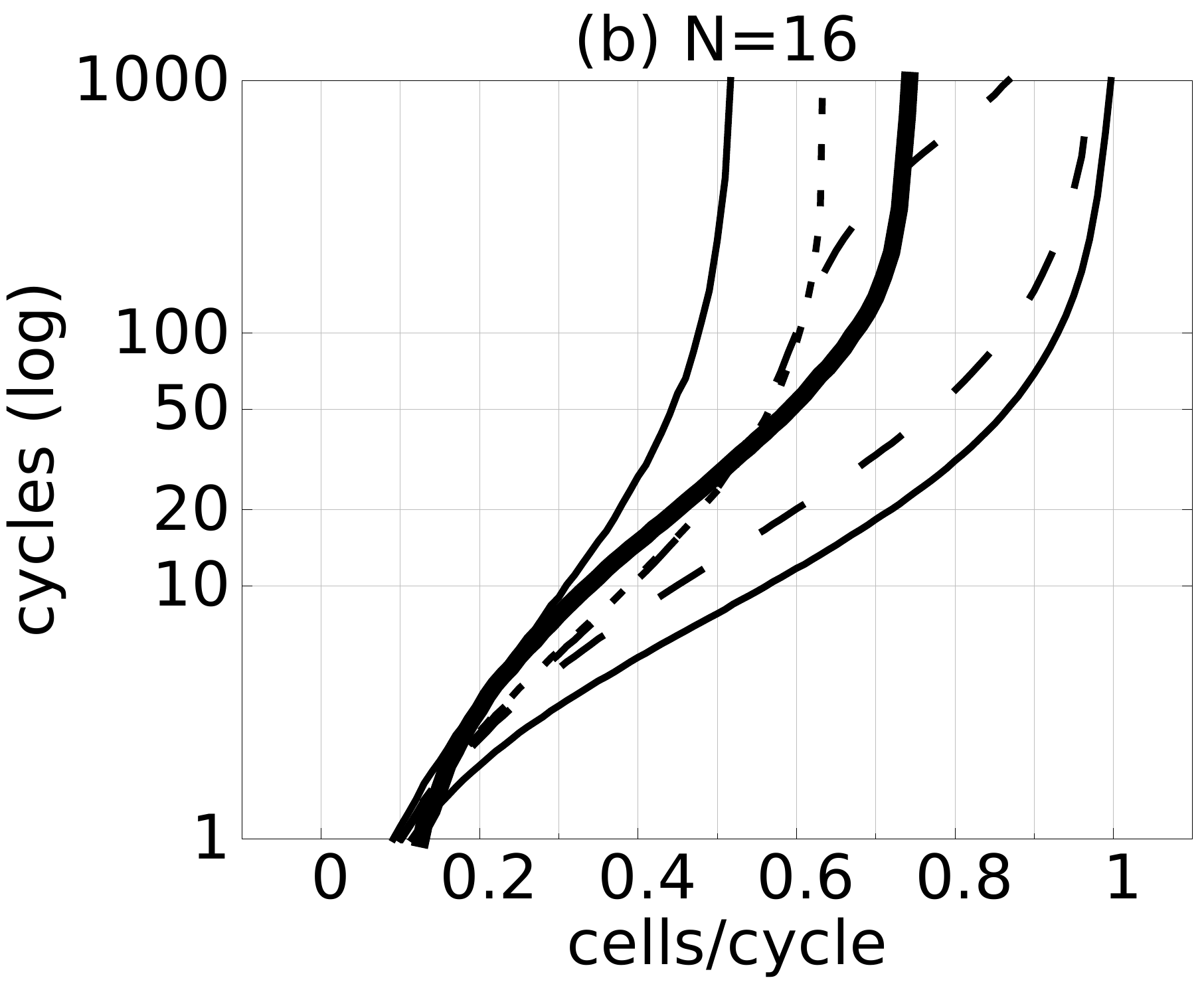}
	\end{tabular}
		\\	
		\ifigpdfFFTH{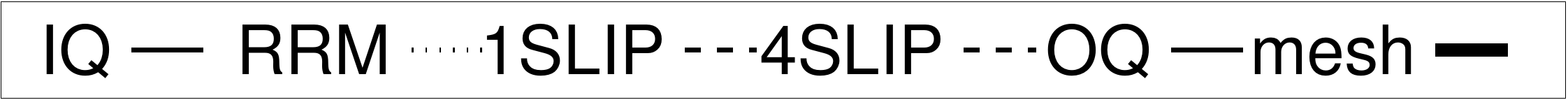}
	\end{tabular}
    \caption{
			Mesh compared to crossbar under synthetic load; links are 4B
			}
    \label{figs:gnuplot:mesh_synth}
\end{figure}

\begin{figure}[t!]
	\centering
	\begin{tabular}{c}
	\begin{tabular}{cc}
		\ifigpdfFFTH{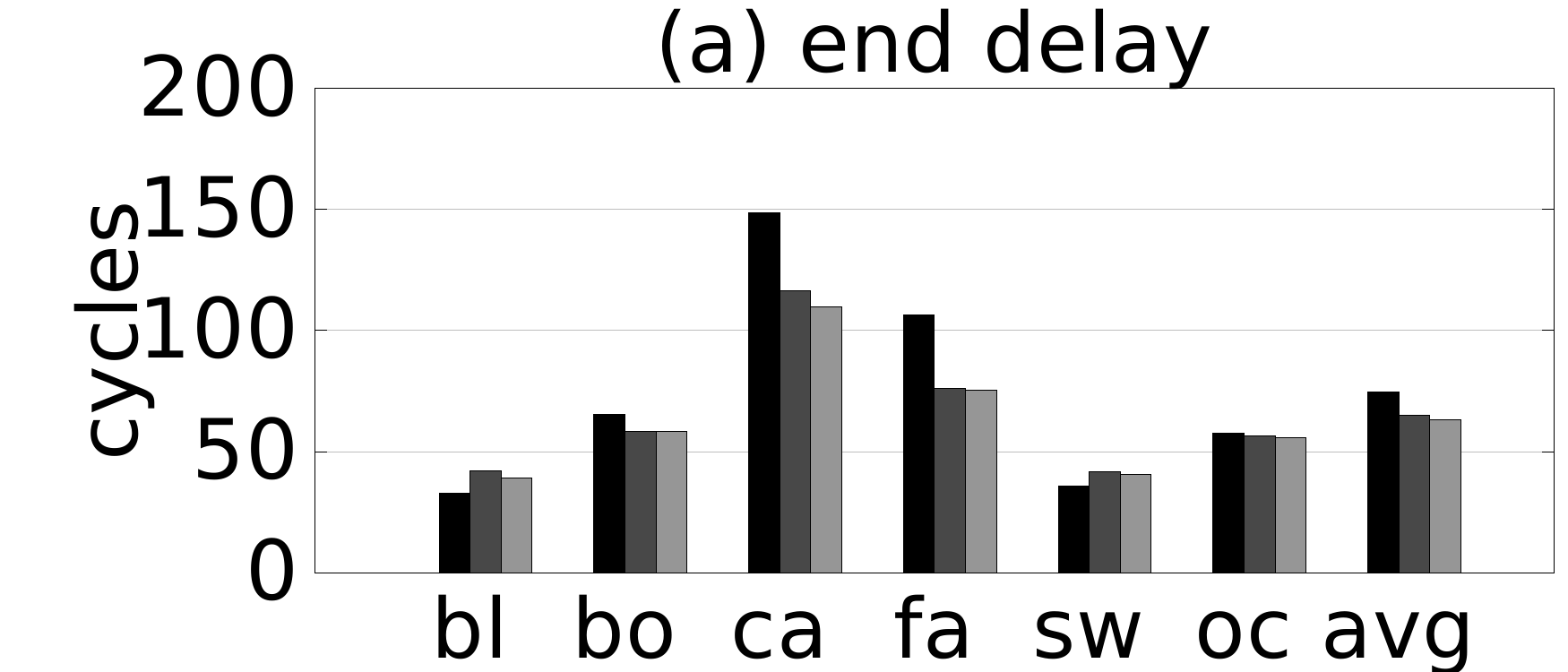}
		&
		\ifigpdfFFTH{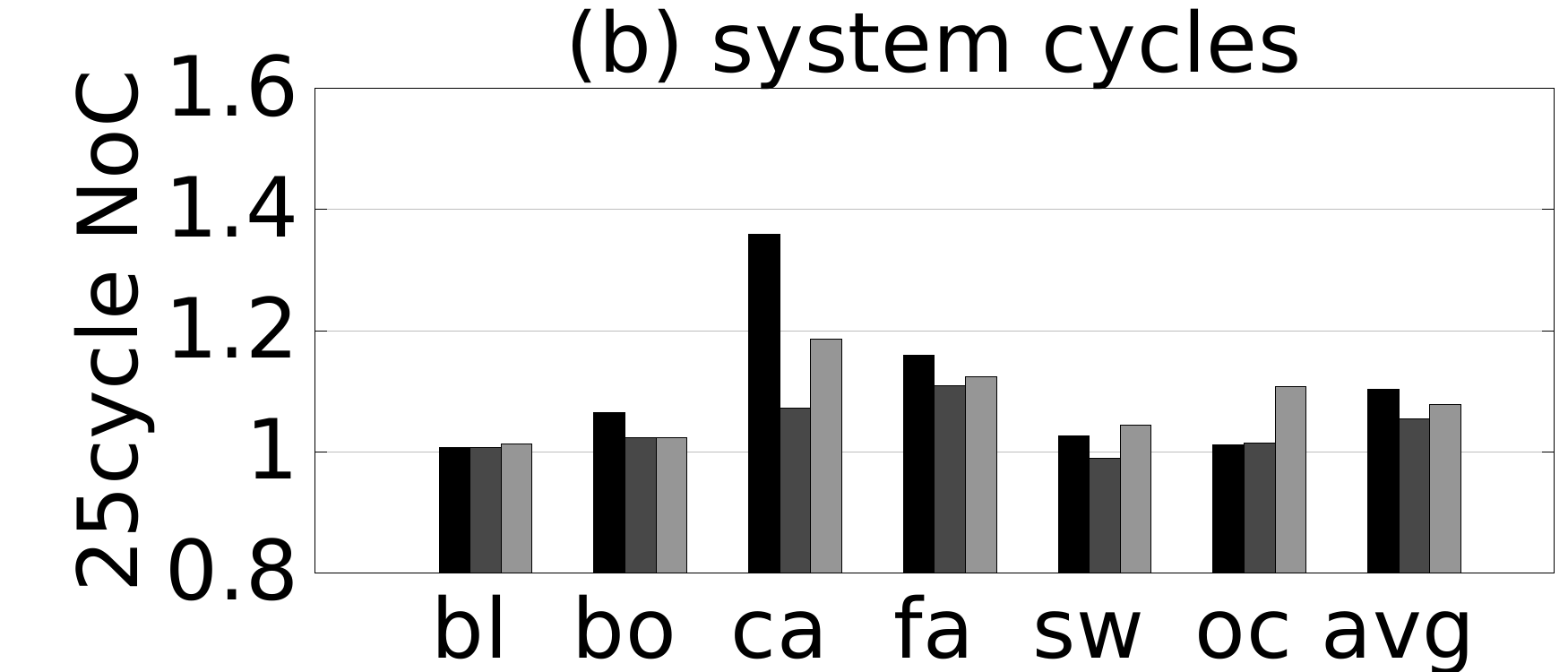}
	\end{tabular}
		\\
		\ifigpdfFFTH{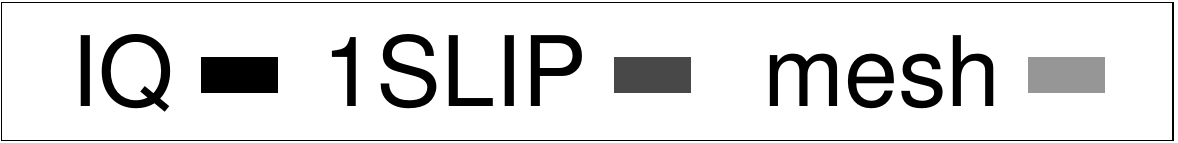}
	\end{tabular}
    \caption{
			Mesh compared to crossbar under benchmark load (execution driven);
			all instances are \numports;
			\streamcluster\ and \radix\ are dropped
			}
    \label{figs:gnuplot:mesh_xbar}
\end{figure}

Fig. \ref{figs:gnuplot:mesh_synth} compares mesh and crossbar queue delay under a synthetic load. The mesh is as good as \SLIPone, shifting toward IQ for larger instances (not shown). Such a nice behavior for small instances should be owing to router buffers that provide a kind of speedup. The performance under benchmark load is plotted in Fig. \ref{figs:gnuplot:mesh_xbar}. The plot intentionally drops \ocean\ and \streamcluster\ to reduce both noise and bias in favor of latency; this work-around could be tolerated considering \ocean\ as a good approximation in place of \streamcluster. The mesh is almost as good as expected. The handicap that appears for \canneal\ and \ocean\ is attributed to cells of control messages suffering exceptionally higher delays. Furthermore, for both the crossbar and the mesh, the benchmark load gives the same delays as double the synthetic load. This discrepancy should be due to spikes being roughly double than bursts by serialization. Unfortunately, this paper had no capacity to study correlation further than spike duration. A complementary study is \cite{cambridge}.

\begin{figure}[t!]
	\centering
	\begin{tabular}{c}
	\begin{tabular}{cc}
		\ifigpdfFFTH{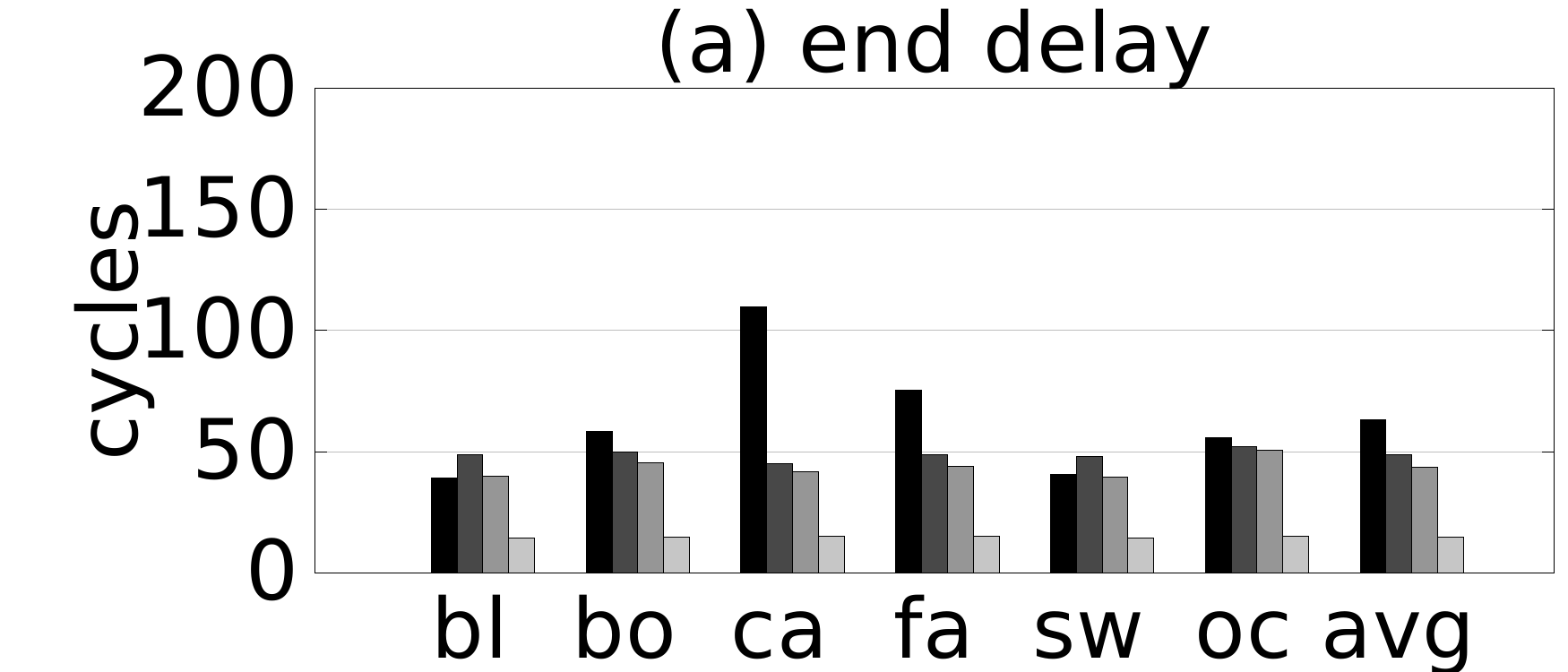}
		&
		\ifigpdfFFTH{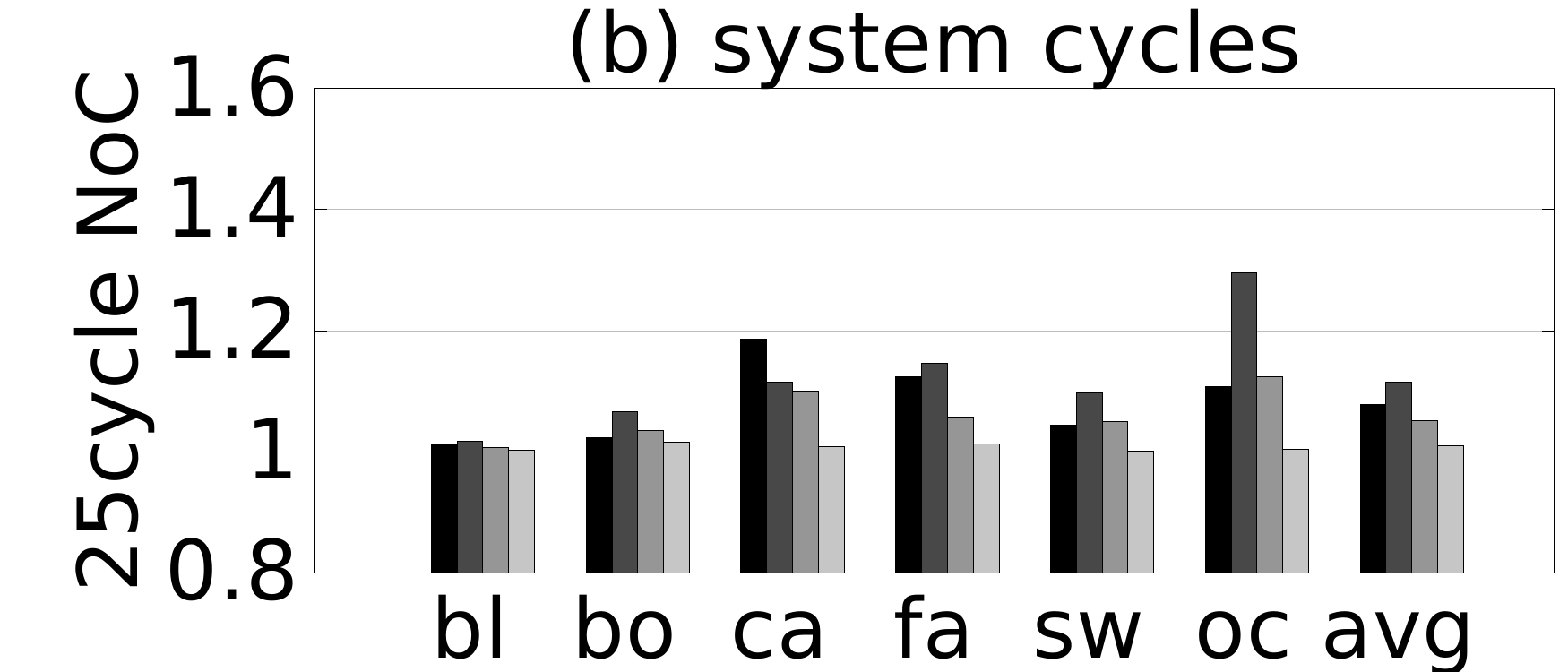}
	\end{tabular}
		\\
		\ifigpdfFFTH{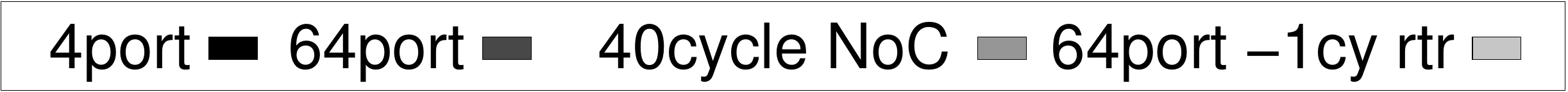}
	\end{tabular}
    \caption{
			Comparing \numports\ and \numportstypical\ meshes; large meshes use either 4-cycle or 1-cycle routers; 
			uniform latency of 40 cycles is also included
			}
    \label{figs:gnuplot:mesh_mesh}
\end{figure}

Fig. \ref{figs:gnuplot:mesh_mesh} compares a \numports\ mesh and a \numportstypical\ mesh. The \numports\ mesh is better, although the end delay is higher by queueing. In particular, the \numportstypical\ mesh degrades performance by a clear \nummeshperflarge. Moreover, performance is worse when compared to a uniform latency of 40 cycles. This is why researchers are proposing techniques to improve locality by optimizing placement of cache blocks \cite{udipi, hybrid}. A third bar in Fig. \ref{figs:gnuplot:mesh_mesh}, corresponding to a \numportstypical\ mesh using one-cycle routers, represents router optimization techniques \cite{mullins_isca, mullins_asia, expressvcs_isca, expressvcs_dac}. Clearly, a large mesh outperforms a small mesh when such techniques are employed. However, the small mesh can be improved by fixing the interleaving of control and data cells, as detailed in Sec. \ref{sec:uber}.

\section{Uber}

\label{sec:uber}

\begin{figure}[t!]
    \centering
    \ifigpdfFFTH{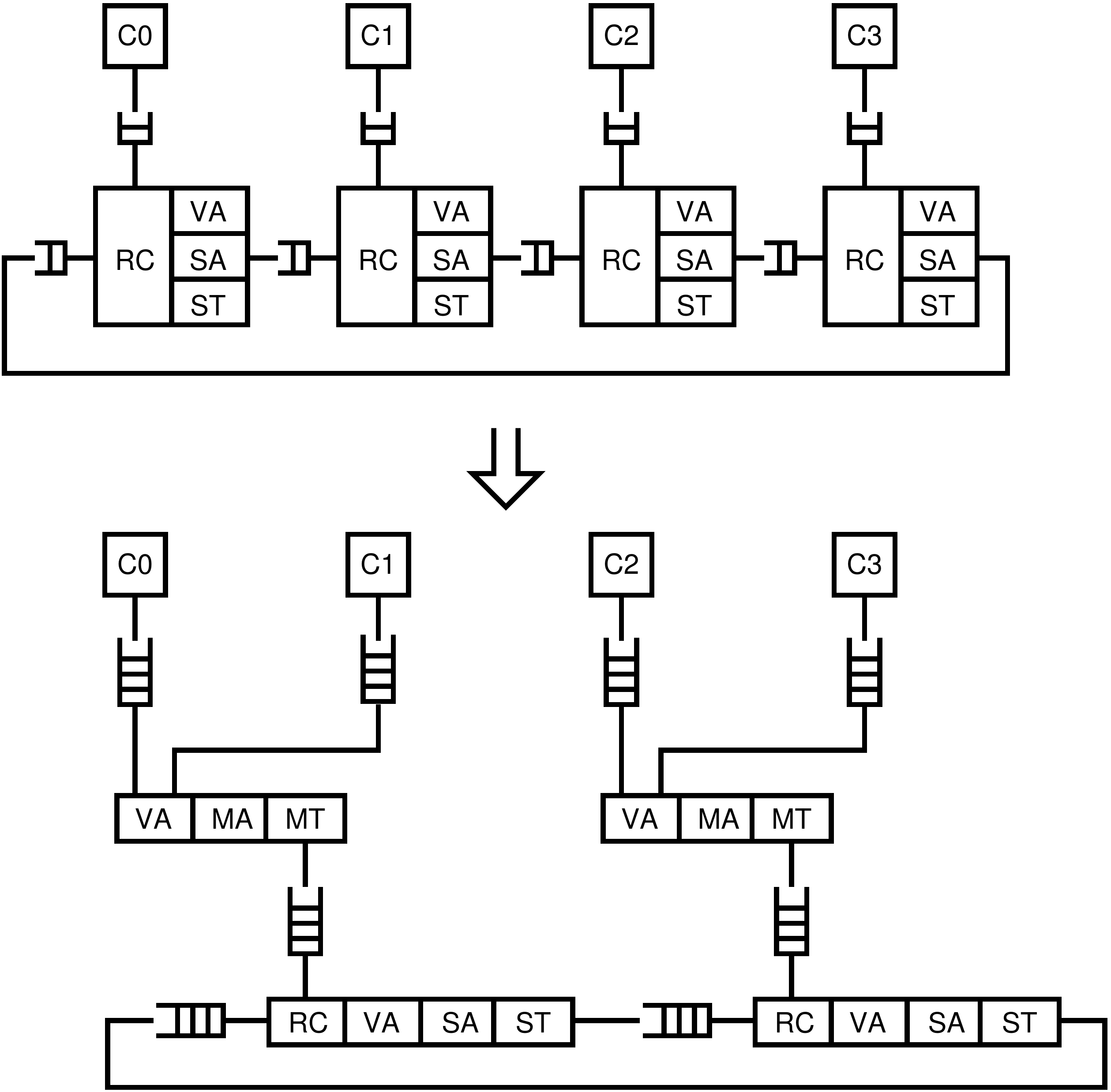}
    \caption{\uber\ schematic}
    \label{figs:xfig:uber_schem}
\end{figure}

Fig. \ref{figs:xfig:uber_schem} gives a schematic for \uber. Typically, techniques rely on bypassing \cite{expressvcs_isca} or speculating router pipelines \cite{mullins_isca}. Set aside the additional hardware, these techniques complicate circuit timing closure \cite{mullins_asia, expressvcs_dac}. As a simpler alternative, \uber\ combines NoC ports. Although combining ports is a well-known practice \cite{ppin}, \uber\ is a novel application on NoCs. Unlike traditional interconnects, where latency is mainly determined by router chip IO \cite{ppin}, the advent of NoCs gives a good opportunity to rethink the role of buffers. Thus, seeing the buffers as the compiler and the router as the processor, \uber\ is a MIPS \cite{mips} equivalent for NoCs.

In an effort to tune the configuration developed in the previous sections into a realistic organization, there are two main steps to take. First, edges implement input in place of output queueing. Extensive evaluations suggest that results are hardly affected on average. Second, buffers are finite, and credit backpressure is added from the switch to the edges and from the edges to the cores. In turn, limited buffers imply deadlocks from protocol dependencies \cite{ppin}. This paper follows the traditional approach to separate control messages (request, forward) and cache blocks (response) in virtual networks. Moreover, virtual networks are strictly prioritized. Preliminary results suggest that virtualization roughly halves end message delay. This differentiation does not change our conclusions. On the contrary, virtualization is a form of QoS for a critical part of the common case. Nevertheless, virtualization indeed relaxes the criticality of HoL blocking by nearly doubling the slack for queueing. 

Thus, this paper predicts a sharply tripolar division of meshes will emerge with hundreds-cores. \uber\ will have a capacity around \numsetuptotalcapscala, will be highly loaded, and will use buffers as a counterweight for simpler routers. Note, however, that even using a small mesh, the NoC bandwidth is likely to be more critical for a particular subset of benchmarks (such as \bodytrack, \canneal, and \facesim) than the whole set. Given that benchmark speedup is hard to sustain \cite{hill_amdhal}, the assumption on linear scaling of bandwidth is aggressive. Dark silicon \cite{darksilicon} could be a second reason. The second pole will have a capacity beyond 16 Tb/s, will be lightly utilized, and will use more complicated router pipelines. And, the third pole will comprise even larger configurations and embarrassingly low utilizations, much like today's NoCs. Bufferless routing \cite{bless}, as well as SMART \cite{smart} is particularly applicable for this third pole. 

Ongoing work will contribute comparisons between \uber\ and bufferless routing \cite{bless}, particularly with respect to energy. Ongoing work also address optimizations for \uber. Updating cache block placement to better balance load by reducing correlation (spikes) is a novel constraint that previous works have not taken into consideration \cite{udipi}.

\section{Conclusion}

Judging from state-of-the-art benchmarks ---where corresponding slack for queueing is large in contrast to latency--- utilizing buffers will play key role in scaling to hundreds-cores. Although \uber\ seems to keep queueing within the nominal slack, more work is needed on optimizations to improve scalability. Future work will also elaborate further by looking more carefully, inside the benchmark boxes. CUDA workloads are also important to analyze \cite{bakhoda}. Either way, by no means has this paper found the ultimate NoC. This paper has contributed an important simplification, as well as a better understanding of stress.

\tiny

\setstretch{1}

\bibliographystyle{abbrv}

\bibliography{references}

\end{document}